\documentclass[12pt]{article}
\usepackage{graphicx}
\usepackage{cite}
\usepackage{vmargin}
\usepackage{amsmath}
\usepackage{amstext}
\usepackage{xspace}

\def\asymerr#1#2#3{\hbox{\rule{0pt}{12pt}${#1}^{+{#2}}_{-{#3}}$}}

\setmarginsrb{1.25in}{1.25in}{1.25in}{1.0in}{0.0in}{0.0in}{0.0in}{0.50in}

\newcommand\pubnumber{DPF2013-80}
\newcommand\pubdate{\today}

\def\princeton{Department of Physics, Princeton University\\
Princeton, NJ 08544, USA}

\def\myemail{\footnote{Email:  mooney@princeton.edu}}
          
\def\Wenu{$\mathrm{W(e\nu)H}$}
\def\Wmunu{$\mathrm{W(\mu\nu)H}$}
\def\Wtaunu{$\mathrm{W(\tau\nu)H}$}
\def\Wlnu{$\mathrm{W(\ell\nu)H}$}
\def\Zee{$\mathrm{Z(ee)H}$}
\def\Zmumu{$\mathrm{Z(\mu\mu)H}$}
\def\Znunu{$\mathrm{Z(\nu\nu)H}$}
\def\Zll{$\mathrm{Z(\ell\ell)H}$}
\def\Hbb{$\mathrm{H} \to \mathrm{b\bar{b}}$}

\newcommand{\ptjj}{\ensuremath{p_{\mathrm{T}}\mathrm{(jj)}}\xspace}

\newcommand{\ptV}{\ensuremath{p_{\mathrm{T}}\mathrm{(V)}}\xspace}

\newcommand{\HERWIG} {{\textsc{herwig}}\xspace}
\newcommand{\MADGRAPH} {\textsc{MadGraph}\xspace}
\newcommand{\POWHEG} {\textsc{powheg}\xspace}

\newcommand{\pt}{\ensuremath{p_{\mathrm{T}}}\xspace}
\newcommand{\GeV}{\ensuremath{\,\text{Ge\hspace{-.08em}V}}\xspace}
\newcommand{\TeV}{\ensuremath{\,\text{Te\hspace{-.08em}V}}\xspace}

\def\Title#1{\begin{center} {\Large #1 } \end{center}}
\def\Author#1{\begin{center}{ \sc #1} \end{center}}
\def\Address#1{\begin{center}{ \it #1} \end{center}}

\newcommand\pubblock{\rightline{\begin{tabular}{l} \pubnumber\\
         \pubdate  \end{tabular}}}
\newenvironment{Abstract}{\begin{quotation}}{\end{quotation}}
\newenvironment{Presented}{\begin{quotation} \begin{center}
                         \end{center}
      \begin{center}\begin{large}}{\end{large}\end{center} \end{quotation}}

\begin{document}
\begin{titlepage}
\pubblock

\vfill
\Title{Search for the SM Higgs Boson Produced in Association with a Vector Boson and Decaying to Bottom Quarks}
\vfill
\Author{ Michael Mooney\myemail \\ \vspace{1mm} \rm On behalf of the CMS Collaboration}
\Address{\princeton}
\vfill
\begin{Abstract}
A search for the Higgs boson produced in association with a W or Z boson and decaying to bottom quarks is presented. A sample of approximately 24 $\mathrm{fb}^{-1}$ of data recorded by the CMS experiment at the Large Hadron Collider, operating at center-of-mass energies of 7 \TeV~and 8 \TeV~in 2011 and 2012, respectively, is used to search for events consistent with the signature of two b jets recoiling with high momentum from a \Wlnu, \Zll, or \Znunu~decay, where $\mathrm{\ell}$ = electron or muon (or hadronically-decaying tau particle in the case of W bosons). Observed signal significance and 95\% confidence level upper limits on the production cross section relative to the Standard Model prediction are presented for the 110-135 \GeV~Higgs mass range.
\end{Abstract}
\vfill
\vspace{0.25in}
\begin{center}PRESENTED AT\end{center}
\begin{Presented}
DPF 2013\\
The Meeting of the American Physical Society\\
Division of Particles and Fields\\
Santa Cruz, California, August 13--17, 2013\\
\end{Presented}
\vfill
\end{titlepage}

\section{Introduction}

The discovery of a Higgs boson \cite{atlas1,cms1} is one of the most important accomplishments of modern experimental particle physics, as in the Standard Model (SM) the Higgs mechanism is considered the explanation for the electroweak symmetry breaking mechanism.  This particle is so far consistent with the SM expectation in terms of observables such as spin, parity, and couplings \cite{atlas2,cms2}.  However, it has not yet been established whether or not this boson decays to fermions, including b quarks.  At the mass of the recently observed Higgs boson, the SM Higgs boson decays predominantly into $\mathrm{b\bar{b}}$ with a branching ratio of roughly 57\%. Therefore, the observation of the \Hbb~decay is essential in determining the nature of the Higgs boson.

At the LHC the main SM Higgs boson production mechanism is gluon-gluon fusion. However, in this production mode, the detection of the \Hbb~decay is considered nearly impossible due to the massive background arising from dijet production expected from quantum-chromodynamic (QCD) interactions. The same holds true to a lesser degree for the next highest cross section production mode, vector-boson fusion (VBF), and for production in association with a $\mathrm{t\bar{t}}$ pair -- though in these processes there is still a handle on QCD rejection using properties of the additional jet activity and/or the presence of isolated leptons. The \Hbb~search with the best handle against backgrounds and highest search sensitivity is the process in which a low-mass Higgs boson is produced in association with a vector boson (W or Z), with the vector boson decaying leptonically.

We report on a search for the Standard Model Higgs boson in the pp $\to$ $\mathrm{VH(b\bar{b})}$ production mode where $\mathrm{V}$ is either a W or a Z boson. The analysis is performed using a data sample corresponding to an integrated luminosity of 5.0 fb$^{-1}$, collected in 2011 by the CMS experiment at a center-of-mass energy of 7 \TeV, and also using a data sample of 19.0 fb$^{-1}$ at 8 \TeV~collected during 2012.  The following final states are included in the search: \Wenu, \Wmunu, \Wtaunu, \Zee, \Zmumu, and \Znunu.

\section{Triggers}

The trigger paths for the \Wenu, \Wmunu, and \Zll~channels consist of several single-lepton and double-lepton triggers with tight lepton identification. Leptons are also required to be isolated from other tracks and calorimeter energy deposits to maintain an acceptable trigger rate. For the \Wmunu~and \Zmumu~channels, for the 2011 (2012) dataset, the trigger thresholds for the muon transverse momentum, \pt, are in the range of 17 (24) to 40 \GeV, where the highest threshold is used without additional isolation requirements.  For the \Wenu~channel, for the 2011 dataset, the electron \pt threshold ranges from 17 to 30 \GeV. The lower-threshold paths require two jets and a minimum requirement on the norm of an online estimate the missing transverse energy vector.  For the 2012 dataset, the single isolated-electron trigger uses a 27 \GeV~threshold.  For the \Zee~channel, dielectron triggers with lower \pt thresholds (17 and 8 \GeV) and tight isolation requirements are used.  For the \Wtaunu~channel trigger, a tau jet from a one-prong hadronically-decaying tau particle is selected for. The \pt of the charged track candidate within the tau jet is required to be above 20 \GeV, and the \pt of the tau jet above 35 \GeV.  An additional requirement of a minimum of 70 \GeV~is placed on the missing transverse energy.

For the \Znunu~channel, combinations of several triggers are used. A trigger with missing transverse energy greater than 150 \GeV~is used for the complete dataset in both 2011 and 2012. During 2011 this trigger was used in conjunction with triggers that require the presence of two central jets ($|\eta|<$ 2.6) with $p_T >$ 20 \GeV~and missing transverse energy thresholds of 80 and 100 \GeV, depending on the luminosity.  During 2012 this trigger was used in conjunction with a trigger that required two central jets above $p_T >$ 30 \GeV~(60 and 25 \GeV~for instantaneous luminosity above 3 $\times 10^{33}$ cm$^{-2}$s$^{-1}$) and a missing transverse energy threshold of 80 \GeV, with additional requirements: at least one pair of jets with \pt greater than 100 \GeV, and no jet with \pt greater than 40 \GeV~closer than 0.5 in azimuthal angle to the missing transverse energy direction.  In order to increase signal acceptance at lower values of missing transverse energy, triggers that require jets to be identified as coming from b quarks are used. For these triggers, two central jets with \pt above 20 or 30 \GeV, depending on the luminosity conditions, are required. It is also required that at least one central jet with \pt above 20 \GeV~be tagged by the online combined secondary vertex (CSV) b-tagging algorithm \cite{btag}.

\section{Analysis}

Below is a brief summary of the analysis strategy employed in the $\mathrm{VH(b\bar{b})}$ search.  A more complete discussion on event reconstruction, event selection, and analysis techniques can be found in \cite{cms3}.

\subsection{Event Selection}

The event selection, largely unchanged since the first result with the full 7 \TeV~dataset \cite{cms4}, is based first on the kinematic reconstruction of the vector bosons and the Higgs boson decay into two b-tagged jets.  Backgrounds are then substantially reduced by requiring a significant boost of the \pt of the vector boson and the Higgs boson \cite{boost}, which tend to recoil away from each other with a large azimuthal opening angle between them.  For each mode, events are split into different categories based on \ptV.  Different signal and background composition in each boost region yields different sensitivity and the signal extraction is performed separately in each category in order to maximize analysis sensitivity -- the final result combines the individual results of the different categories and modes.  These categories, labeled as ``low", ``intermediate", and ``high" boost, are defined differently for each mode.  For \Wmunu~and \Wenu~the three categories are 100-130 \GeV, 130-180 \GeV, and 180+ \GeV.  For \Znunu~the ranges are similar:  100-130 \GeV, 130-170 \GeV, and 170+ \GeV.  In the case of \Zll~there are only two categories, defined with ranges of 50-100 \GeV~and 100+ \GeV.  Note that there is only one category for \Wtaunu, 120+ \GeV, as a result of limitations in the number of Monte Carlo simulation events post-selection.

Reconstruction of the vector boson (W or Z) is done by selecting isolated, central leptons and/or high missing transverse energy.  Orthogonality is ensured between the different modes when selecting the vector boston candidate.  The reconstruction of the \Hbb~decay is made by requiring the presence of two central ($|\eta| <$ 2.5) jets above a minimum \pt threshold.  If more than two such jets are found in the event, the pair of jets with the highest total dijet transverse momentum, \ptjj, is selected.  The two jets are then required to be tagged by the CSV algorithm, satisfying a minimum threshold of the CSV discriminator that varies between modes.

\subsection{b-jet Energy Regression}

The mass resolution for the two b jets from the Higgs decay is approximately 10\% with a small bias on the mass of a few percent. The mass resolution and bias are improved by applying a b-jet energy regression technique.  A specialized boosted decision tree (BDT) regression algorithm (performed with gradient boosting) is trained on simulated signal \Hbb~events with inputs that include detailed information about the jet structure and that help differentiate jets from b quarks from light-flavor jets. These include variables containing information about several properties of a secondary vertex when present, information about tracks, charged constituents, and other variables related to the energy reconstruction of the jet.  If a non-isolated soft lepton is also present within the jet (arising from semileptonic B decays), then properties of this lepton is used in the regression.

The resulting improvement on the $\mathrm{H(b\bar{b})}$ mass resolution is approximately 15-20\% for the \Zll~modes and 7-12\% for the other modes.  The additional gains for \Zll~arise from the use of missing transverse energy in the regression, as there is no neutrino production from vector boson decay in these modes only.  Applying this technique results in an increase in the analysis sensitivity of roughly 15\%.

\subsection{Background Control Samples}

Backgrounds arise from production of W and Z bosons in association with jets (from all quark flavors), single and pair-produced top quarks, dibosons and QCD multijet processes.  Control samples (CS's) are defined to isolate and study these backgrounds.  Suitable CS's are identified in data and used to correct the Monte Carlo yield estimates in the signal region for several of the most important background processes, including production of W and Z bosons in association with jets (one CS for two light jets, another for at least one heavy flavor jet) and $\mathrm{t\bar{t}}$ production.  This leads to five CS's for \Znunu~(in which both W+jets and Z+jets backgrounds significantly contribute) and three CS's for all other modes.

A set of simultaneous fits is then performed to the distributions of discriminating variables in the control regions, separately in each channel, to obtain consistent scale factors by which the Monte Carlo yields in the signal region are adjusted.  Note that in this fit, the W+jets and Z+jets backgrounds are split based upon how many b jets are associated with the Higgs candidate (zero, one, or two) at the level of MC truth.

\subsection{Signal Extraction}

In the final stage of the analysis, and to better separate signal from background under different Higgs boson mass hypotheses, a BDT classifier algorithm is trained separately at each mass value using simulated samples for signal and background that pass the event selection described above. The most discriminating input variables make use of additional jet activity, the mass of the reconstructed $\mathrm{H(b\bar{b})}$ candidate, and b-tagging information.  The shape of the output distribution of this BDT algorithm is the final discriminant with which a binned maximum-likelihood fit is performed to search for events resulting from Higgs boson production.  A rebinning technique is applied to each BDT distribution before being fit in order to ensure sufficient Monte Carlo statistics in each bin.

\begin{figure} 
\includegraphics[width=.49\textwidth]{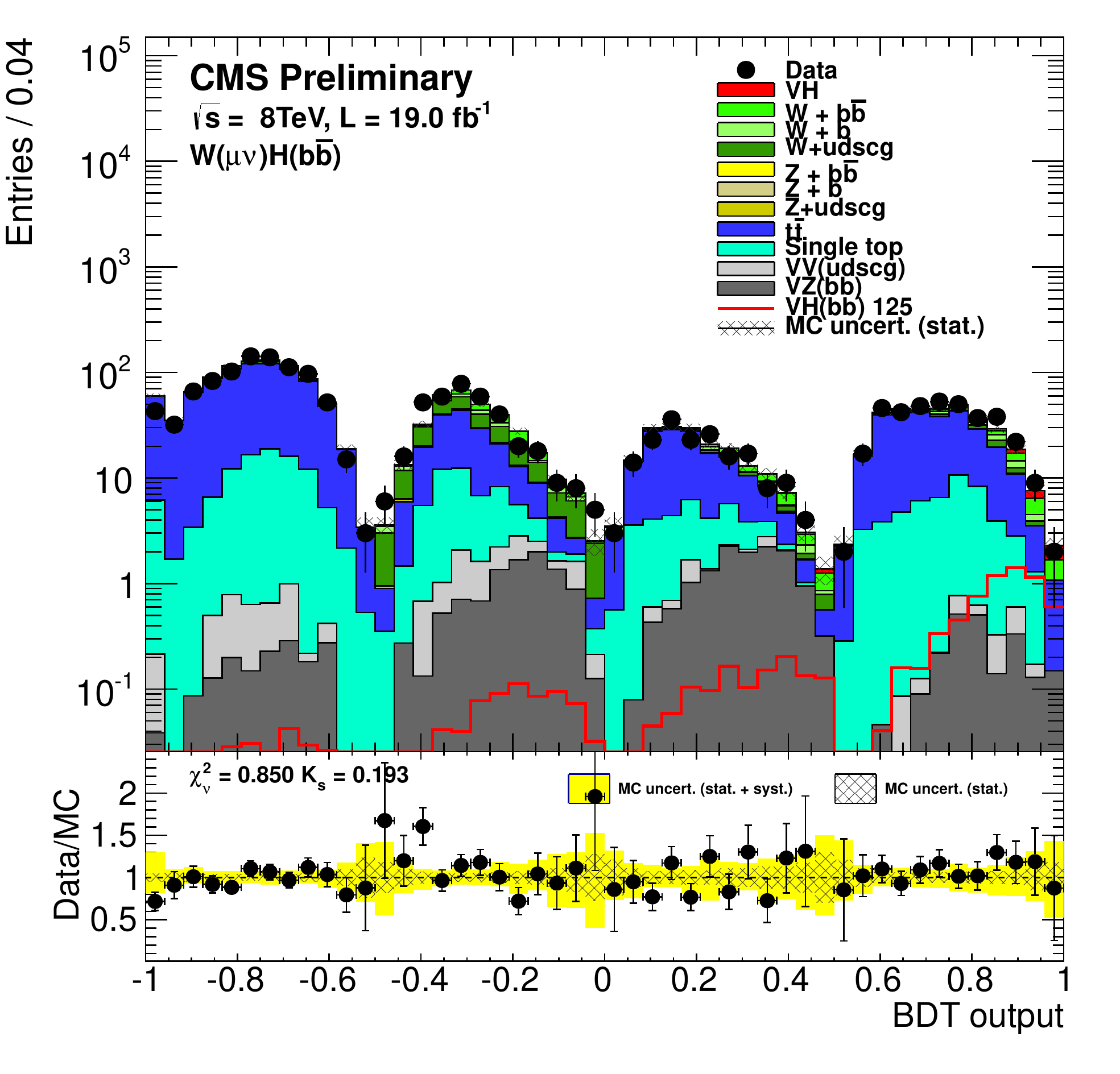} 
\includegraphics[width=.49\textwidth]{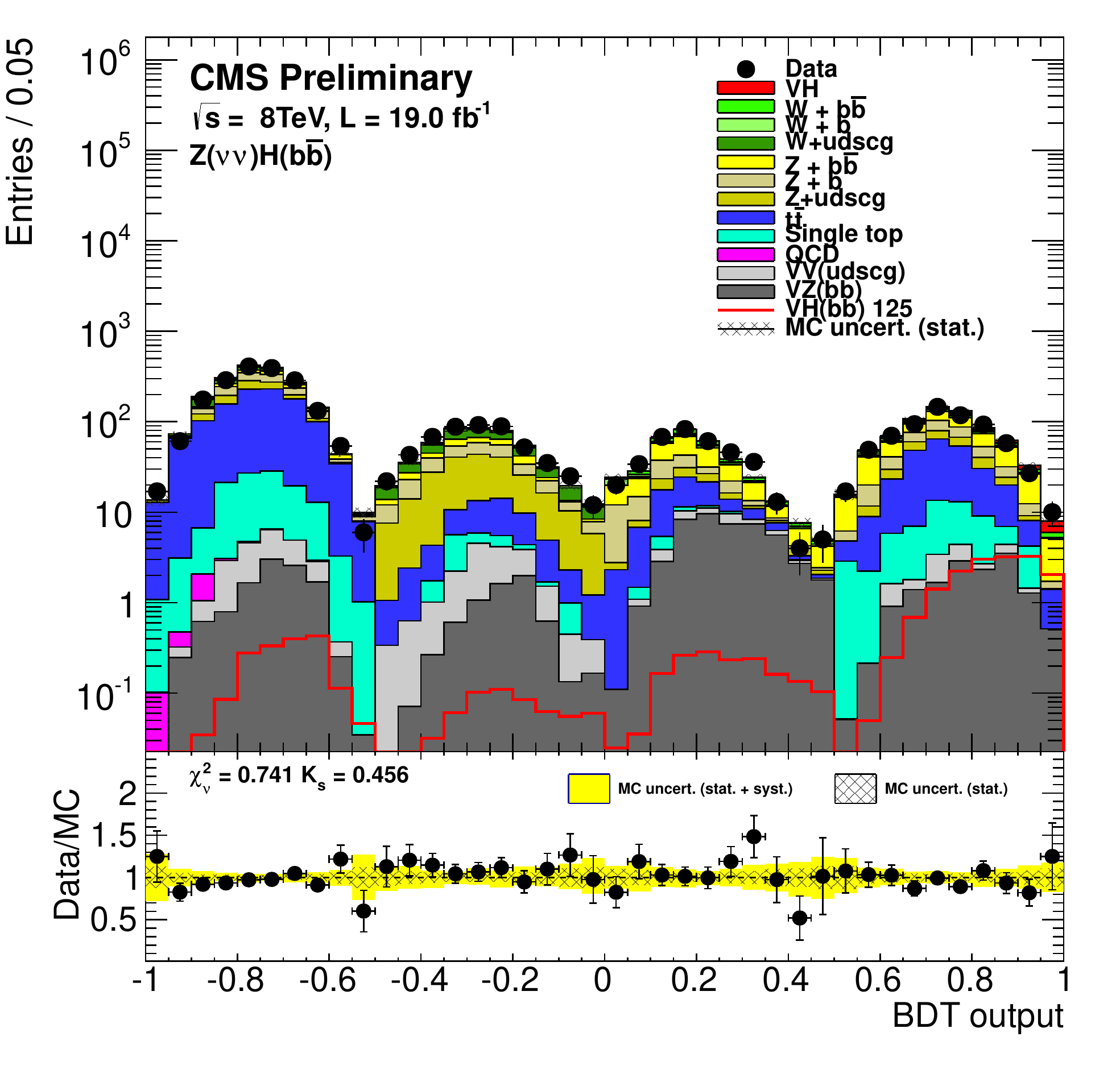} 
\caption{Multi-BDT output distributions for \Wmunu~(left) and \Znunu~(right) in the high \ptV bin, for data (points with errors), all backgrounds, and signal, after all selection criteria have been applied.} 
\label{fig1}
\end{figure}

Additionally, in the \Wlnu~and \Znunu~modes (except for \Wtaunu), a ``multi-BDT" technique is employed to maximize search sensitivity.  This technique, used only in the 8 \TeV~analysis, divides the sample into four distinct subsets that are enriched in $\mathrm{t\bar{t}}$, V+jets, dibosons, and VH, which are then fit simultaneously.  The subsets are defined via cuts on the output from three additional background-specific BDT classifier algorithms that train signal against one background at a time.  The addition of this technique results in an increase of approximately 10\% in sensitivity for the relevant modes.  Figure~\ref{fig1} shows some examples of the multi-BDT distributions for the 8 \TeV~data. 

\subsection{Systematic Uncertainties}

The dominant systematic uncertainties due to detector effects come from the b-tagging simulation and the jet energy scale and resolution.  Data/MC b-tagging scale factors are measured in heavy-flavor-enhanced samples of jets that contain muons and are applied consistently to jets in signal and background events.  The associated uncertainty is on the order of 3-15\% depending on the channel and the specific process.  The jet energy scale is varied within one standard deviation as a function of jet \pt and $\eta$, giving a 2-3\% yield variation. The effect of the uncertainty on the jet energy resolution is evaluated by smearing the jet energies according to the measured uncertainty, giving a 3-6\% variation in yields.  These uncertainties are propagated to both the shape and normalization of the BDT distributions used in the final fit.  Uncertainties from the background yield scale factors obtained from the CS fits are also included.

The total VH signal cross section has been calculated to NNLO order accuracy, with a total uncertainty of 4\%. This analysis is performed in the boosted regime, and thus, potential differences in the \pt spectrum of the V and H between data and Monte Carlo generators could introduce systematic effects in the signal acceptance and efficiency estimates.  Both NLO electroweak \cite{EWK1,EWK2,EWK3} and NNLO QCD \cite{QCD} corrections are applied to the signal samples, with an uncertainty of 2\% (5\%) for the electroweak (QCD) corrections for both ZH and WH modes.  For the V+jets backgrounds, the difference between the BDT shapes using the \MADGRAPH and \HERWIG Monte Carlo generators is taken as a systematic uncertainty, while for the $\mathrm{t\bar{t}}$ background the same is done using the \MADGRAPH and \POWHEG Monte Carlo generators.  Uncertainties of 15\% are used for both single top and diboson background yields, consistent with previous CMS measurements \cite{ST,VV}.

\section{Results}

\begin{figure} 
\includegraphics[width=.48\textwidth]{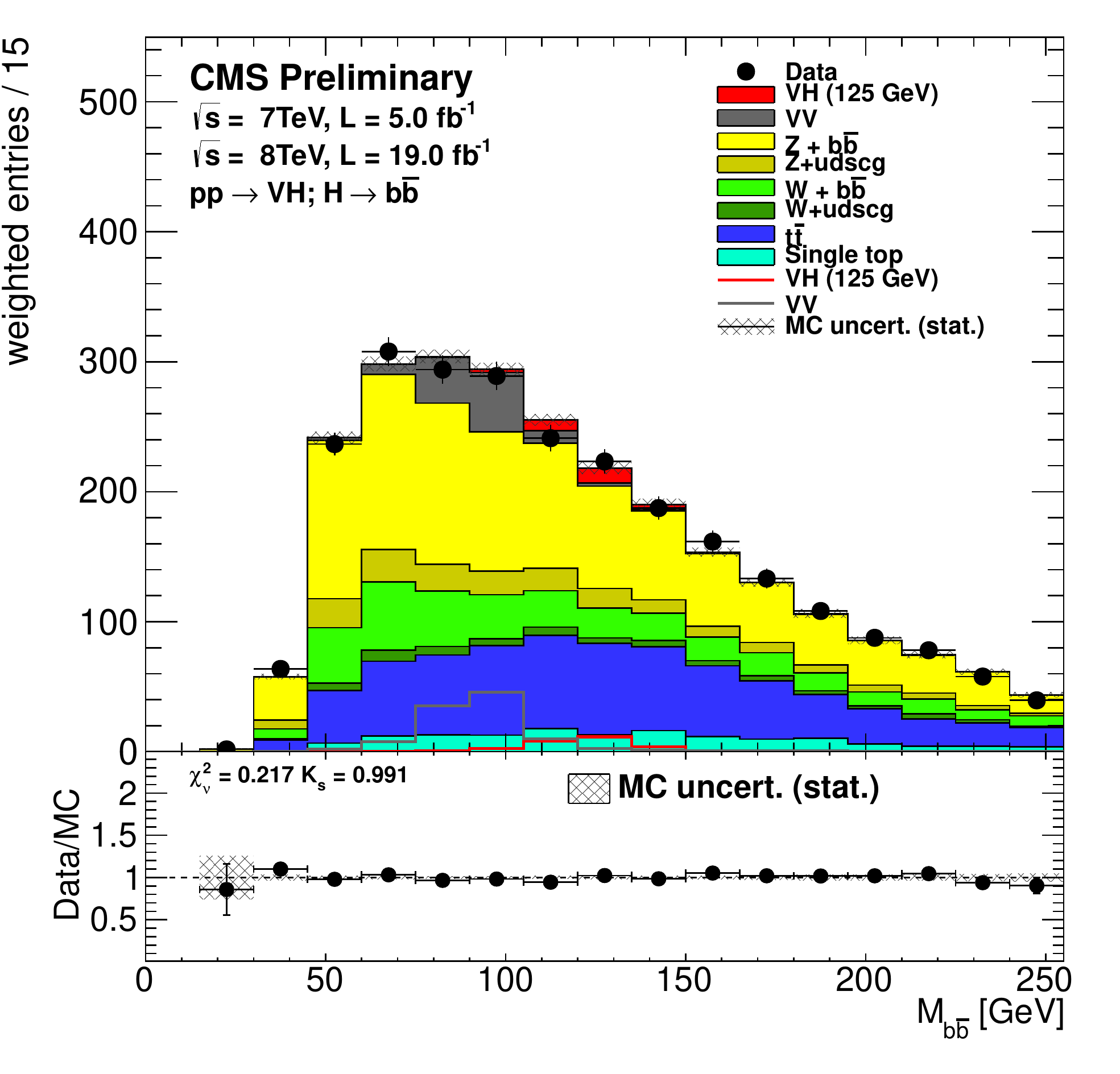} 
\includegraphics[width=.48\textwidth]{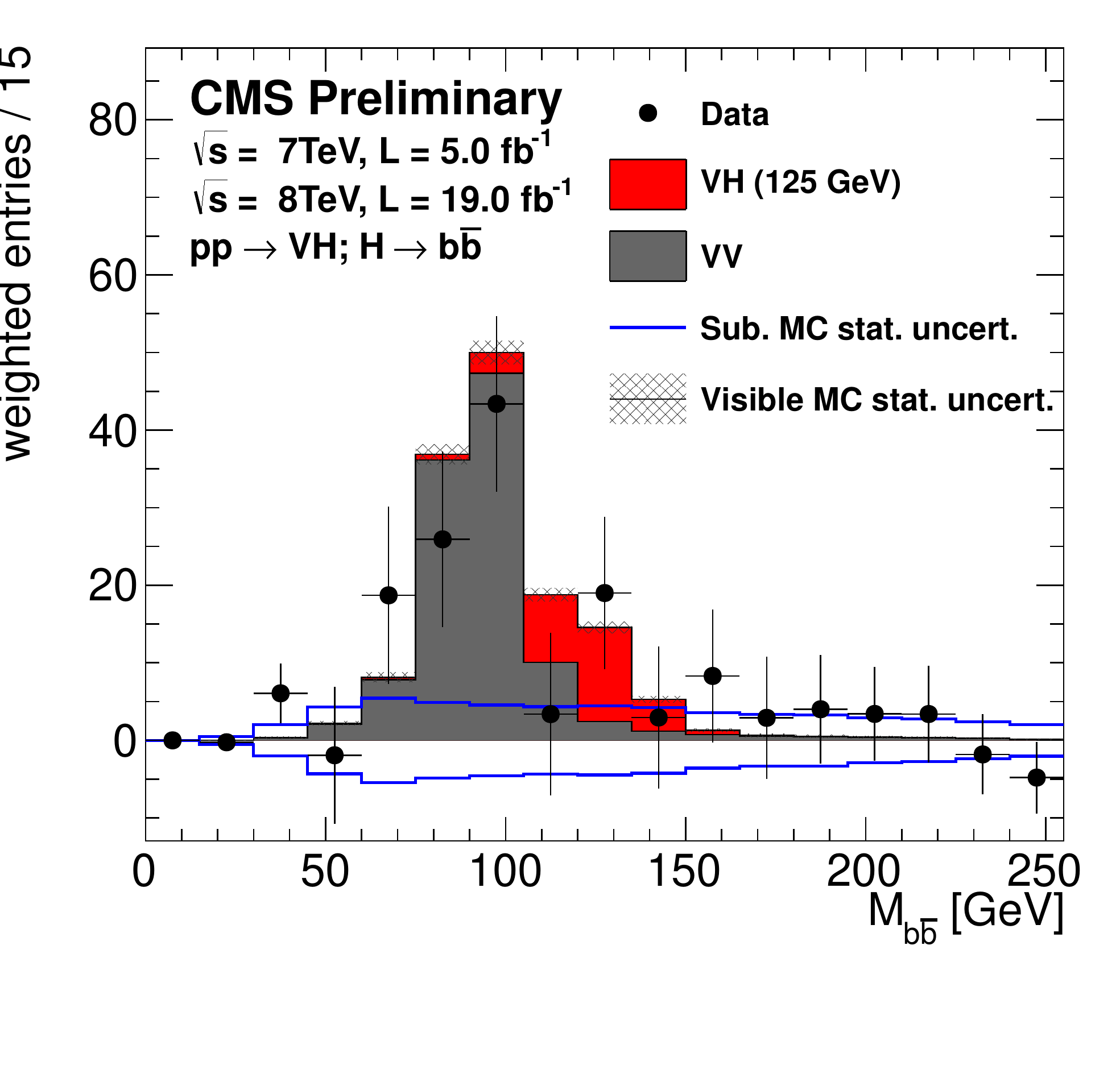} 
\caption{Combined signal region M(jj) distribution weighted by S/(S+B) for each mode/category (left), and background-subtracted combined M(jj) distribution showing the diboson background and signal excess (right).} 
\label{fig2}
\end{figure}

The combined signal region M(jj) distribution, weighted by S/(S+B) for each mode and category, is shown in Fig.~\ref{fig2} along with the background-subtracted distribution (keeping the diboson background out of the subtraction).  Upper limits at the 95\% CL on the pp $\to$ VH production cross section are obtained for Higgs boson masses in the 110-135 \GeV~range.  These limits result from fitting the BDT (and multi-BDT) shapes described above for the presence of a Higgs boson signal above what is expected from all background components.

\begin{figure} 
\includegraphics[width=.48\textwidth]{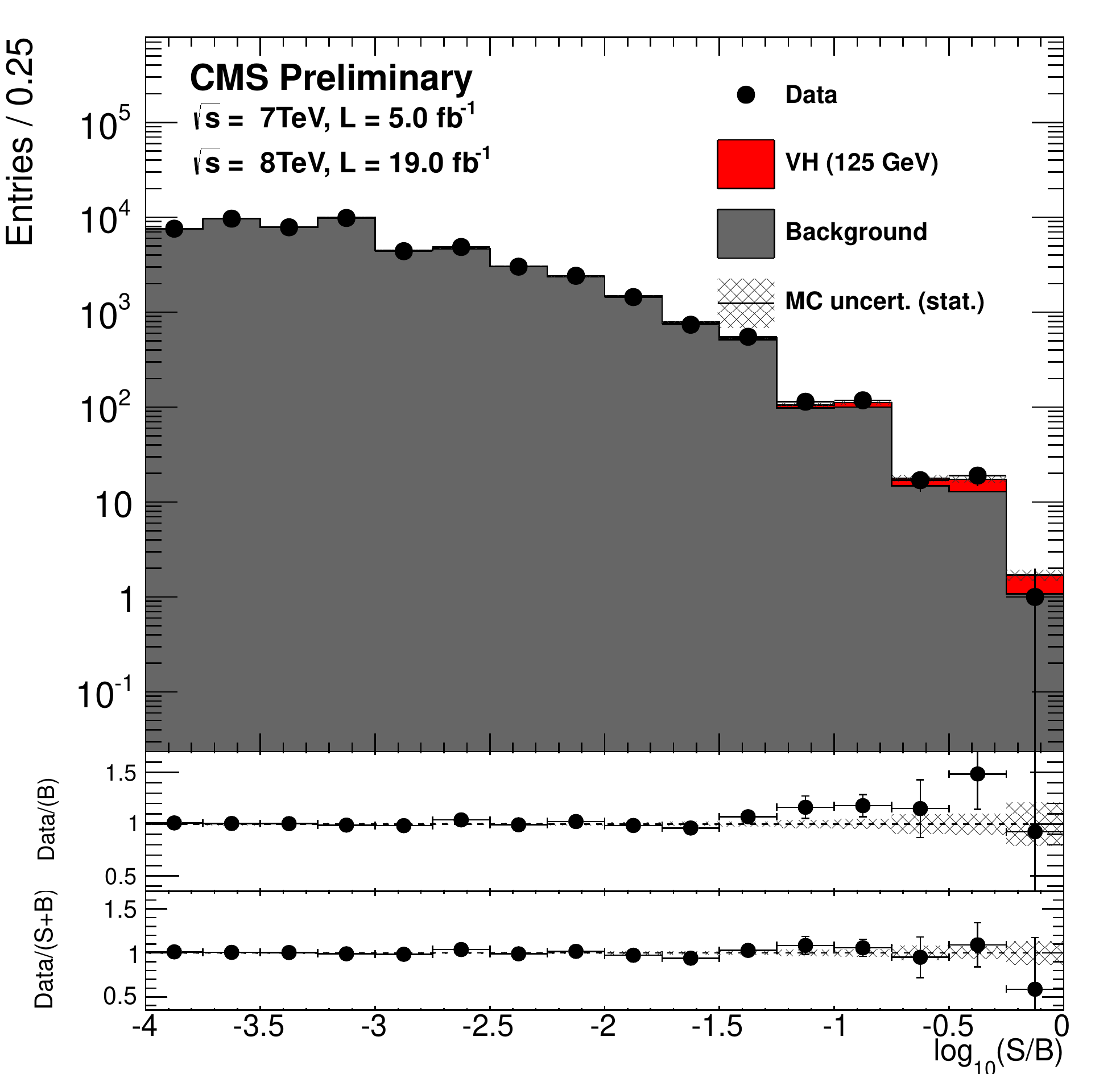} 
\includegraphics[width=.48\textwidth]{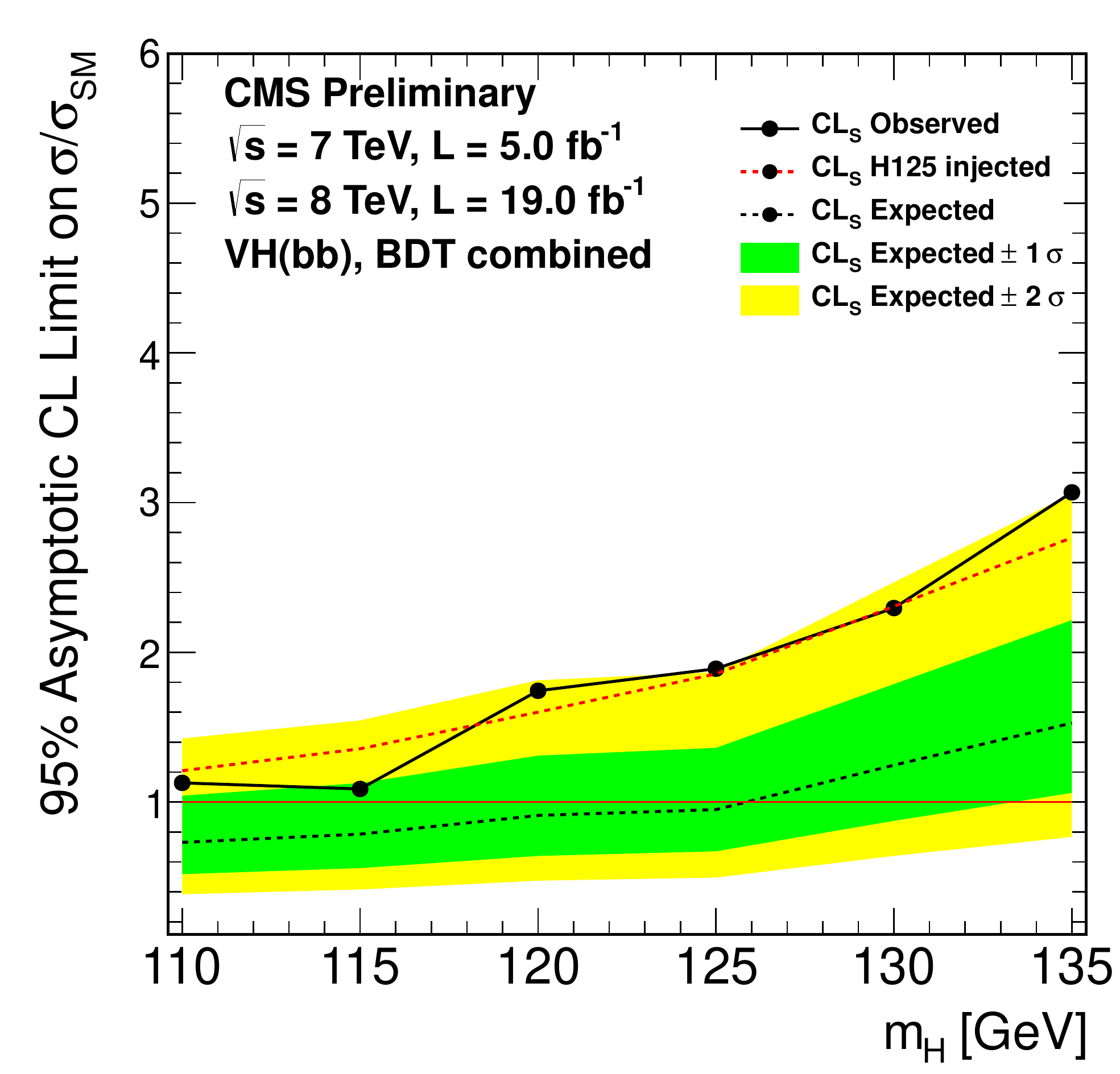} 
\caption{Combination of all BDT fit results (signal and background) as a function of $\log{\mathrm{(S/B)}}$ of each individual bin in the separate BDT fits (left), and expected/observed 95\% CL upper limits on the product of the Higgs production cross section times the \Hbb~branching ratio with respect to the expectations for a Standard Model Higgs boson (right).} 
\label{fig3}
\end{figure}

Presented in Fig.~\ref{fig3} is the combination of all BDT fit results.  The observed limits at each mass point, the median expected limits, and the 1$\sigma$ and 2$\sigma$ bands are calculated using the modified frequentist method CLs and displayed in Fig.~\ref{fig3} and Table~\ref{tab1}.  At a Higgs mass of 125 \GeV, an excess of events is observed above the expected background with a local significance of 2.1 standard deviations, which is consistent with the expectation from the production of the SM Higgs boson.  The signal strength corresponding to this excess, relative to that of the SM Higgs boson, is $1.0 \pm 0.5$.

As a validation of the multivariate technique, BDT discriminants are trained using the diboson sample as signal, and all other processes, including VH production (at a mass of 125 \GeV), as background for 8 \TeV~data only.  The observed local signal significance is greater than 7 standard deviations with a signal strength of \asymerr{1.19}{0.28}{0.23}.

\begin{table}[htbp]
\caption{Expected and observed $95\%$ CL upper limits on the
product of the VH production cross section times the \Hbb~branching ratio with respect to the expectations for a Standard Model Higgs boson.}
\label{tab1}
\begin{center}
{\small
\begin{tabular}{cccccccc} \hline\hline
m(H)\hspace{0.5mm}[\GeV]   &   110 & 115    & 120   & 125   & 130 & 135\\  \hline
Exp. Limit         &  0.73  & 0.79  & 0.91  & 0.95  & 1.25  & 1.53\\
Obs. Limit         &  1.13  & 1.09  & 1.74  & 1.89  & 2.30  & 3.07\\
\hline\hline
\end{tabular}
}
\end{center}
\end{table}

\end{document}